\documentstyle[12pt]{article}

\newcommand{\sect}[1]{\setcounter{equation}{0}\section{#1}\indent}

\begin{document}

\topmargin 0pt
\oddsidemargin 5mm
\def\bbox{{\,\lower0.9pt\vbox{\hrule \hbox{\vrule height 0.2 cm
\hskip 0.2 cm \vrule height 0.2 cm}\hrule}\,}}
%%%%%%%%%%%%%%%%%%%%%%%%%%%%%%%%%%%%%%%%%%%%%%%%
\def\a{\alpha}
\def\b{\beta}
\def\g{\gamma}
\def\G{\Gamma}
\def\d{\delta}
\def\D{\Delta}
\def\e{\epsilon}
\def\ve{\varepsilon}
\def\z{\zeta}
\def\t{\theta}
\def\vt{\vartheta}
\def\r{\rho}
\def\vr{\varrho}
\def\k{\kappa}
\def\l{\lambda}
\def\L{\Lambda}
\def\m{\mu}
\def\n{\nu}
\def\o{\omega}
\def\O{\Omega}
\def\s{\sigma}
\def\vs{\varsigma}
\def\S{\Sigma}
\def\vphi{\varphi}
\def\av#1{\langle#1\rangle}
\def\pa{\partial}
\def\na{\nabla}
\def\hg{\hat g}
\def\un{\underline}
\def\ov{\overline}
\def\cF{{{\cal F}_2}}
\def\Hsl{H \hskip-8pt /}
\def\Fsl{F \hskip-6pt /}
\def\cFsl{\cF \hskip-5pt /}
\def\ksl{k \hskip-6pt /}
\def\pasl{\pa \hskip-6pt /}
\def\tr{{\rm tr}}
\def\tcF{{\tilde{{\cal F}_2}}}
\def\tg{{\tilde g}}
\def\cLNSI{{\cal L}^{\rm NS}_{\rm I}}
\def\bcLNSI{{\bar\cLNSI}}
\def\shalf{\frac{1}{2}}
\def\nn{\nonumber \\}
\def\w{\wedge}
%%%%%%%%%%%%%%%%%%%%%%%%%%%

\def\cmp#1{{\it Comm. Math. Phys.} {\bf #1}}
\def\cqg#1{{\it Class. Quantum Grav.} {\bf #1}}
\def\pl#1{{\it Phys. Lett.} {\bf B#1}}
\def\prl#1{{\it Phys. Rev. Lett.} {\bf #1}}
\def\prd#1{{\it Phys. Rev.} {\bf D#1}}
\def\prr#1{{\it Phys. Rev.} {\bf #1}}
\def\prb#1{{\it Phys. Rev.} {\bf B#1}}
\def\np#1{{\it Nucl. Phys.} {\bf B#1}}
\def\ncim#1{{\it Nuovo Cimento} {\bf #1}}
\def\jmp#1{{\it J. Math. Phys.} {\bf #1}}
\def\aam#1{{\it Adv. Appl. Math.} {\bf #1}}
\def\mpl#1{{\it Mod. Phys. Lett.} {\bf A#1}}
\def\ijmp#1{{\it Int. J. Mod. Phys.} {\bf A#1}}
\def\prep#1{{\it Phys. Rep.} {\bf #1C}}

%%%%%%%%%%%%%%%%%%%%%%%%%%%%%

\begin{titlepage}
\setcounter{page}{0}

\begin{flushright}
COLO-HEP-382 \\
hep-th/9705139 \\
May 1997
\end{flushright}

\vspace{5 mm}
\begin{center}
{\large Coupling of branes and normalization of effective actions in
string/M-theory }
\vspace{10 mm}

{\large S. P. de Alwis\footnote{e-mail:  
dealwis@gopika.colorado.edu}}\\
{\em Department of Physics, Box 390,
University of Colorado, Boulder, CO 80309}\\
\vspace{5 mm}
\end{center}
\vspace{10 mm}

\centerline{{\bf{Abstract}}}
We discuss issues involving p (D)- brane charge quantization and the
normalization of effective actions,
in string/M-theory. We also construct the action  of  (the bosonic  
sector of)
eleven dimensional supergravity in the presence of two and five  
branes  and
discuss  (perturbative) anomaly cancellation.

\end{titlepage}
\newpage
\renewcommand{\thefootnote}{\arabic{footnote}}
\setcounter{footnote}{0}

\setcounter{equation}{0}

\sect{Introduction}
A major claim of string theory is that it has no parameters. All of  
the
fundamental experimentally measureable constants should in principle  
be
determined by the dynamics. This means in particular that the string  
coupling,
which is the expectation value of the scalar dilaton field, as well  
as the size
and shape of the six-dimensional
compact space, should be fixed by some (hitherto unknown) mechanism.  
Clearly
these values are not determined in perturbation theory and it is  
therefore
important to understand the non-perturbative dynamics of strings.
The discovery of various strong-weak coupling relations (S-dualities)  
in string
theory, as well as
the connection of string theory to M-theory \cite{ew1}, leads one to  
hope that
one may be on the threshold of understanding
the non-perturbative dynamics of string theory. Coupled with the  
insight coming
from the incorporation of D-branes\cite{jp1} into string  
theory\footnote{For
reviews of D-branes with extensive references to the original  
litereature see
\cite{jprev} and for  M-theory see \cite{ptrev}.} it seems that one  
can fix
(non-perturbatively) the ten dimensional gauge and gravitational   
constants
in terms of the dilaton expectation value and the string scale.

Some work on these issues was done by this author in reference  
\cite{sda1}.
Part of the motivation for this paper is to elaborate on those  
arguments,
particularly in relation to
the claim that these constants are fixed non-perturbatively at the  
string
scale.  The
idea is that these constants are the exact values that go into the   
low energy
string effective action (obtained by integrating out the sub-string  
scale
fluctuations) which is used as a starting point for compactification  
and
renormalization group evolution down to low energies. The paper also  
contains a
detailed discussion of the coupling of two and
five-branes to background supergravity (with two-branes that may end  
on
five-branes) in an explicit manner.
In section two (which borrows heavily from \cite{duffrev}) we discuss  
solutions
of string effective
actions with p-form gauge fields. The object of the discussion is to  
show that
purely from the demand for the existence of solutions of these  
actions, and a
natural identification of the metric
on the brane, one gets the dilaton dependence for the action for the  
zero modes
 that is expected for D-branes from lowest order  perturbation theory
calculations \cite{jprev} (when the gauge fields are of the R-R  
type). One also
gets the canonical form \cite{ew1} of the R-R action. As a  
consequence we argue
that the
dilaton dependence coming from lowest order perturbation theory is  
probably
valid non-perturbatively. In section three, much of which is a review  
of
earlier work, we use the Dirac quantization relations for p-branes,   
T-duality,
and type IIB $SL (2,Z)$, to show how the tensions of  all D-branes,  
and also of
the two and five branes of M-theory, can be expressed in terms of
the fundamental length scale. In addition we show that the  
ten-dimensional
gravitational coupling
of string theory and the eleven dimensional gravitational coupling of  
M-theory,
are determined
in terms of the string scale. We subject these relations to various  
consistency
checks. Again the
main point is that these relations are exactly (non-perturbatively)  
valid at
the string scale. While these relations have been presented before
\cite{jp1},\cite{sda1} the detailed discussion given here will  
hopefully
further clarify the issues involved. In section four we present  
arguments to
the effect that the form of the gauge field dependence in the D-brane  
action is
also valid non-perturbatively. In section five we extend our  
discussion to
theories which may be phenomenologically relevant. In  section six we  
discuss
the coupling of two- and five-branes in M-theory (with two-branes  
which may sit
on five-branes \cite{pt},\cite{as}) to
the low energy M-theory background. In section seven we use this  
formalism to
discuss (perturbative) anomaly cancellation in  the presence of  
five-branes and
in particular suggest
a possible resolution of a puzzle  pointed out recently  \cite{ew4}.

\sect{Solutions of effective actions}
The bosonic part of the closed superstring action is given by
\begin{equation}\label{string}
S=-T\int_{W_2}d^2\xi
{\shalf}\sqrt{-\g}\g^{ij}\pa_iX^M\pa_jX^NG_{MN}-T\int_{W_2}P(B_2)-{1\o 
ver
4\pi}\int_{W_2}\phi R^{(2)}.
\end{equation}
 The tension   $T={1\over 2\pi\a '}$, where $l_s\equiv\sqrt\a '$ is  
the
associated string (length) scale, is a physical quantity which is  
measurable in
principle on the assumption that strings can exist as isolated  
objects i.e.
like electrons and unlike quarks.\footnote{This is assumed to be the  
case even
though the string coupling is  probably fixed at some value of order  
one at the
string scale (rather than $10^{-2}$ as in QED).}
The above action describes a string (whose world sheet is $W_2$)  in  
a
background consisting of a condensate of its massless bosonic  
fluctuations,
giving the metric ($G$) the NS-NS two form field ($B$) \footnote{  
$P(B)={1\over
2!}B_{MN}{\pa X^M\over\pa\xi^i}{X^N\over\pa\xi^j}d\xi^Id\xi^j$ is the  
pull back
of the target space field to the world sheet. Our convention for  
n-forms is
$A_n={1\over n!}A_{i_1,\ldots i_n}dx^{i_1}\w\ldots\w  
dx^{i_n},~F_{n+1}=dA_n
={1\over n!}\pa_{i_0}A_{i_1,\ldots i_n}dx^{i_0}\w dx^{i_1}\w\ldots\w
dx^{i_n}$.}and  the dilaton scalar field ($\phi$). The string  
coupling is given
by
\begin{equation}\label{}
g = <e^{\phi}>_0.
\end{equation}
The coupling of the theory is not a free parameter but is determined
dynamically. In   perturbative string theory this value is actually
undetermined and one has to appeal to non-perturbative effects, which  
are
certainly far from being understood at this point, in order to fix  
it. Later on
we will give two different identifications of this value, which are  
consistent
with each other and T-duality,  in terms of ratios of physical  
lengths.

The quantum consistency of this world sheet action imposes the  
requirement that

\begin{equation}\label{}
T\int_{C_2} H_3=2\pi n,~~~n\e{\cal Z},
\end{equation}
where locally $H_3=dB$, and this fixes the normalization of  $B$ in  
terms of
the string tension.
Also as long as the world sheet is closed (no open-closed string  
interactions)
the action (\ref{string}) is invariant under the guage transformation
$B\rightarrow B+d\L$.
Now unlike  a particle, a string can only propagate in backgrounds  
which
satisfy the requirement of conformal invariance, and this leads in a  
well-known
fashion to $\b$-function equations that
are effective equations of motion for the background fields. These  
equations
imply the existence of an effective action for the condensed massless  
fields
that takes the form (keeping only the leading terms in an $\a '$  
expansion)

\begin{equation}\label{effaction}
I=-{1\over 2\k^2}\int_{M_{10}}\sqrt{-G}e^{-2\phi}\left\{  
R-4(\nabla\phi
)^2+{1\over12}H^2 \right\}.
\end{equation}
It is important to note  that the relative coefficients in this  
action are
fixed by world sheet conformal invariance. Also, although the  
Lagrangian
density apears with an overall dilaton factor appropriate to a  
spherical world
sheet, since the beta function equations are independent of global  
topology one
expects this action to be valid to all orders in perturbation
theory\footnote{There are  terms which may arise due to the so-called
Fischler-Susskind effect, coming from degenerate Riemann surfaces,  
but these
are not expected to modify the terms above.}, and is perhaps even an  
exact
statement of the  interactions of these fields at low energies. For  
the rest of
this paper we will assume that the latter is actually the case.

The target space metric $G$ that occurs in the above formulae is  
called the
string metric since
it is the metric in which (upon elimination of the intrinsic metric  
in
(\ref{string})  the string action is given by the area
of the world sheet; i.e. as $\int_{W_2} \sqrt{-G}$ \footnote{Wherever  
there is
no confusion we will use the same letter for the field defined on  
$M_{10}$ and
the pulled-back field on a p-brane world volume $W_{p+1}$.}. From
(\ref{effaction}) we see that the ten dimensional gravitational  
constant is
given as $ G_N^{10}={\k^2 g^2\over 8\pi}$. Alternatively one can  
transform to
the so-called
Einstein frame in which the effective gravitational action has the  
standard
form. Thus $G_{\mu\nu}^E=e^{{-\shalf}\phi}G_{\mu\nu}$. In this case  
the
gravitational constant is $ G_N^{10}={\k^2 \over 8\pi}$ but the  
string tension
\footnote{In this paper $2\pi\a '$ will always be the inverse string  
tension in
the string frame.}  is now $T={g^{\shalf}\over 2\pi\a '}$. The  
dimensionless
quantity $G_NT^4 $ is of course independent of the metric.

An action for a p-brane coupling to a background (p+1)-form field may  
be
written as,
\begin{equation}\label{braneaction}
S_p=-T_p\int_{W_{p+1}}d^{p+1}\xi e^{c\phi }\sqrt{\det
G_{ij}}-T_p\int_{W_{p+1}}A_{p+1}.
\end{equation}
Where $G_{ij}$ is the metric on $M_{10}$ pulled back to the world  
volume and in
the last integral the pullback map to the world volume is understood.  
$c$ in
the above is a constant and the field $\phi$ need not be identified  
a'priori
with the dilaton introduced earlier. In general it may be a function  
of all
scalar (composite) fields, but in the subsequent analysis the  
identification
with the standard dialaton will be justified and so we will use the  
same letter
here. The only assumption here is that the first term is given by a  
positive
function of the dilaton times the volume form.
It is convenient to introduce a current (p+1)-form whose dual is a  
closed delta
function (9-p) form
which restricts an integral over $M_{10}$ to $W_{p+1}$.
\begin{equation}\label{delta}
*J_{p+1}=\d_{9-p}(M_{10}\rightarrow W_{p+1}),~~~d*J = d\d=0
\end{equation}
Note that the last equation is simply a current conservation  
condition. Also
this is normalized such that $\int_{M_{9-p}}\d=1$. If for instance we  
take the
p-brane to be embedded in the sub-space spanned by the first p+1  
coordinates
then an explicit expression for this is

\begin{equation}\label{exdelta}
*J_{p+1}=\d_{9-p} = \int_{W_{p+1}}dX^1(\xi )\w...\w  
dX^{p+1}\d^{10}(x-X(\xi
))dx^{p+2}\w...\w dx^{10}.
\end{equation}
The topological term in the above action may then be written as

\begin{equation}\label{}
\int_{W_{p+1}}A_{p+1}=\int_{M_{10}}A_{p+1}\w\d=\int_{M_{10}}A_{p+1}\w  
*J.
\end{equation}
Let us now include a kinetic term  ${1\over 2\k^2}\int  
e^{-b\phi}\shalf
F_{p+2}\w *F_{p+2}$, where $F=dA_{p+1}$, in the low energy effective  
action
(\ref{effaction}). We couple the p-brane to the background by  
replacing the
action $I$ by $I+S_p$. The equation of motion for
$A$ then gives us the generalized Maxwell equation

\begin{equation}\label{}
{1\over 2\k^2}d(e^{-b\phi}*F)=T_p*J,
\end{equation}
whose integrated form is the electric charge equation (Gauss Law),

\begin{equation}\label{electric}
{1\over 2\k^2}\int_{\pa M_{9-p}}e^{-b\phi}*F_{p+2}=T_p.
\end{equation}
By a standard argument the quantum consistency of the p-brane action  
gives,
\begin{equation}\label{magnetic}
T_p\int_{W_{p+2}}F_{p+2}=2\pi n, ~~~n\e{\cal Z}.
\end{equation}
Now we may introduce the dual field strength
\begin{equation}\label{dual}
\tilde F_{8-p}=e^{-b\phi}*F_{p+2}
\end{equation}
By demanding the quantum consistency of the coupling
$T_{6-p}\int_{W_{7-p}}A_{7-p}$ to a (6-p)-brane, we have the  
well-known
generalization of the Dirac quantization condition \cite{rn}

\begin{equation}\label{dirac}
2\pi n =T_{6-p}\int_{W_{8-p}}\tilde F_{8-p}=2\k^2T_{6-p}T_p,
\end{equation}
where the last step follows from (\ref{dual}) and  (\ref{electric}).

Let us now look at the low-energy effective action for  
dilaton-gravity and a
(p+2) form field strength in the  Einstein metric.

\begin{equation}\label{}
I={1\over 2\k^2}\int_{M_{10}} d^{10}x\sqrt{-G^E}(R-\shalf (\nabla\phi
)^2)-{1\over 2\k^2}\int_{M_{10}}e^{-a\phi}\shalf F_{p+2}\w *F_{p+2},
\end{equation}
where the curvature and scalar products are all defined in the  
Einstein metric.
Let us now define the p-brane metric as $G^{(p)}_{MN}=e^{a\phi  
/(p+1)}
G^E_{MN}$. Transforming to this metric we find,
\begin{equation}\label{}
I={1\over 2\k^2}\int_{M_{10}} d^{10}x\sqrt{-G^{(p)}}e^{-4a\phi\over
p+1}(R-\shalf (\nabla\phi )^2)-{1\over  
2\k^2}\int_{M_{10}}e^{-4a\phi\over
p+1}\shalf F_{p+2}\w *F_{p+2}.
\end{equation}
Thus in this metric, the effective low energy action scales under  
constant
shifts of the dilaton,
in the same way that the dilaton gravity action with the tree-form  
field
strength term, scales in the string metric.
It is natural to define the p-brane action in this metric  
as\footnote{An
alternative argument is given in \cite{duffrev}.},

\begin{equation}\label{}
S_p=-T_p\int_{W_p}d^{p+1}\xi \sqrt{\det  
G^{(p)}_{ij}}-T_p\int_{W_p}A_{p+1}.
\end{equation}
 Now let us require that there exist  asymptotically flat solutions  
to the
equations of motion coming from the action $I+S_p$. Then, as shown in  
the
review by Duff et al \cite{duffrev} (see equation (3.38)) this  
determines
\begin{equation}\label{a}
a=\pm (p-3)/2.
\end{equation}
Let us  take the lower sign in (\ref{a}). Transforming to the  
`string' metric
by putting
\begin{equation}\label{}
G^{(p)}_{MN}=e^{-{(p-3)\phi\over 2(p+1)}}G^E_{MN}=e^{-{p-1\over
p+1}\phi}G_{MN}.
\end{equation}
 we have
\begin{equation}\label{}
I={1\over 2\k^2}\int_{M_{10}}\sqrt{-G}e^{-2\phi}\left\{  
R+4(\nabla\phi )^2
\right\}-{1\over 2\k^2}\int_{M_{10}}e^{(p-3)\phi}\shalf F_{p+2}\w  
*F_{p+2}
\end{equation}
for the effective background action
and
 \begin{equation}\label{}
S_p=-T_p\int_{W_p}d^{p+1}\xi e^{-{p-1\over 2}\phi}\sqrt{\det
G_{ij}}-T_p\int_{W_p}A_{p+1}
\end{equation}

For $p=1$, $I$ becomes  the NS-NS sector of the string effective  
action and $S$
becomes the string action in the string metric. This justifies the  
name for
this metric as well as the identification of $\phi$ with the standard
perturbative dilaton. For $p=5$ we have the dual form of the NS-NS  
string
effective action and the action $S_5$ has the characteristic  
$e^{-2\phi}$
dilaton dependence of  a solitonic action.

Let us now choose the upper sign and then transform back to the  
string metric
using
\begin{equation}\label{}
G^{(p)}_{MN}=e^{(p-3)\phi\over 2(p+1)}G^E_{MN}=e^{-{2\phi\over  
p+1}}G_{MN}.
\end{equation}
 The low energy effective action then takes the form,

\begin{equation}\label{}
I={1\over 2\k^2}\int_{M_{10}}\sqrt{-G}e^{-2\phi}\left\{  
R+4(\nabla\phi )^2
\right\}-{1\over 2\k^2}\int_{M_{10}}\shalf F_{p+2}\w *F_{p+2}.
\end{equation}
This is in the form that one expects for the kinetic energy terms of  
the
Ramond-Ramond fields. Transforming  the p-brane action to  the string  
metric we
get,

\begin{equation}\label{}
S_p=-T_p\int_{W_{p+1}}d^{p+1}\xi e^{-\phi}\sqrt{\det
G_{ij}}-T_p\int_{W_{p+1}}A_{p+1}.
\end{equation}
This has exactly the form of the D-brane action obtained from string
perturbation theory (albeit with the gauge field on the brane set to  
zero).

What is remarkable here is that this result has been obtained without  
any use
of string perturbation theory or indeed of the existence of open  
strings. The
point is that the low energy
effective action is an quantum exact action and all that is assumed  
is the
existence of
the relevant p-brane configurations together with a very natural  
assumption
about their actions. Thus the dilaton dependence of the D-action is  
probably
independent of perturbation theory.

\sect{T-duality arguments and D-brane tensions}
We will first briefly review some arguments in  given in  \cite{sda1}  
and
elsewhere\footnote{See the first paper of \cite{jprev} for a review  
of the
relevant literature.}.

Let us  consider asymptotically flat configurations with  
$G_{MN}\rightarrow
\eta_{MN}$,
$\phi\rightarrow \phi_0=const.$ with $e^{\phi_0}=g$ the string  
coupling
constant. The physical tension (energy per unit spatial volume) is  
then
\begin{equation}\label{}
\tau_p=g^{-1}T_p.
\end{equation}
Note that $T_p$ is in fact the electric charge associated with the  
brane and so
the above is a charge-mass relation that follows from  the equality  
of the two
coefficients in the two terms of  (\ref{braneaction}), which in turn  
is a
consequence of super ($\kappa$)-symmetry. Thus it is  in fact a BPS  
formula
\cite{azc}. Now consider configurations with one (spatial) isometry.  
Choosing
coordinates such that the Killing vector is $\pa/\pa x^{10}$ and  
taking that
direction to  to be a circle of radius $R$, T-duality is the  
statement  that
the transformation

\begin{equation}\label{}
R\rightarrow R'={\a '\over R},~~~e^{\phi}\rightarrow e^{\phi  
'}={\sqrt\a '\over
R}e^{\phi},
\end{equation}
 gives us a theory with the same physics (with in particular  the  
same
effective action).
The requirement that the D-brane physical mass remains the same  and  
the
argument  that the p-brane is mapped into a (p$\pm$1) brane under  
T-duality,
then
gives the relation \begin{equation}\label{tensionrecursion}
T_{p-1}=2\pi\sqrt{\a '}T_{p}.
\end{equation}
Now SL(2,Z) duality of type IIB string theory tells us that the  
D-string charge
is equal to the
fundamental string charge (tension) i.e. $T_1 = T={1\over 2\pi\a '}$.  
Thus we
have from (\ref{tensionrecursion}) the formula,
\begin{equation}\label{ptension}
T_p={1\over (2\pi )^p(\sqrt{\a '})^{p+1}}.
\end{equation}
Using also the Dirac quantization formula with $n=1$  
(\ref{dirac})\footnote{It
should be noted that for RR fields and D-branes, we put $b=0$  in
(\ref{electric}) and the curvature and scalar products are taken in  
the string
metric.}, we get
\begin{equation}\label{kappa}
2\k^2=(2\pi )^7\a '^4.
\end{equation}
Thus the quantization rule just serves to fix the relation between  
the ten
dimensional  gravitational constant and the string scale (in the  
string metric)
once the ambiguity arising from the ability to do constant shifts in  
the
dilaton is fixed by our choice of the T-duality rules.

Now let us consider the relationship to M-theory.
The bosonic part of the low energy M-theory effective action is given  
by

\begin{equation}\label{mnormal}
I_{11}=-{1\over  2\k_{11}^2}\int_{M}\sqrt{-G}R-{1\over
2\k_{11}^2}\shalf\int_{M}K_4\wedge *K_4-{1\over  2\k_{11}^2}{1\over
6}\int_{M}C_3\wedge K_4\wedge K_4,
\end{equation}
where $K_4$ is a closed 4-form field strength which may be locally  
written as
$K_4=dC_3$ and the integration is over an 11 dimensional Minkowski  
signature
manifold.
Let us consider 11D configurations which have an isometry generated  
by a
Killing vector $k=\pa/\pa y$ with the standard identification
\begin{equation}\label{yidentify}
y \leftrightarrow y+2\pi\sqrt{\a '}.
\end{equation}
 The Kaluza-Klein ansatz for reducing this to the type IIA low-energy  
effective
action\footnote{see \cite{ptrev}, \cite{duffrev} and references  
therein.} is

\begin{equation}\label{kk1}
ds^2=e^{-{2\over 3}\phi_A  
(x)}G_{\mu\nu}(x)dx^{\mu}dx^{\nu}+e^{{4\over 3}\phi_A
(x)}(dy-C_{\mu}(x)dx^{\mu})^2,
\end{equation}
and

\begin{eqnarray}\label{kk2}
C_3&=&{1\over 3!}C_{MNP}dx^M\w dx^N\w dx^P={1\over  
3!}A_{\mu\nu\l}dx^{\mu}\w
dx^{\nu}\w dx^{\l}+{1\over 2!}B_{\mu\nu}dx^{\mu}\w dx^{\nu}\w dy\nn
&=&A_3+B_2\w dy.
\end{eqnarray}
At large 10 D distances we can then identify the radius of the  
eleventh
dimension (parametrized by $y$) with the (IIA) coupling constant by

\begin{equation}\label{}
R_{11}=\sqrt{\a '}<e^{2\phi_A\over 3}>=\sqrt{\a '}g_A^{2\over 3}.
\end{equation}
This gives a physical definition to the IIA coupling.
By using (\ref{kk1}) and (\ref{kk2}) in (\ref{mnormal}) one obtains  
the IIA
action \cite{duffrev},

\begin{eqnarray}\label{twoa}
I_{IIA}&=&-{1\over 2\k^2}\int_{M_{10}}\sqrt{-G}e^{-2\phi}\left\{  
R-4(\nabla\phi
)^2+{1\over12}H^2 \right\}\nn
& &-{1\over 2\k^2}\int_{M_{10}}(\shalf F_2\w *F_2+\shalf F_4\w  
*F_4)-{1\over
2\k^2}\shalf\int_{M_{10}}F'_4\w F'_4\w B_2,
\end{eqnarray}
where
\begin{equation}\label{mtwoa}
2\k_{11}^2=(2\pi\sqrt{\a '})2\k^2,~ H_3=dB_2,~F_2=dC_1,~F'_4=dC_3,~
F_4=F'_4-H_3\w C_1.
\end{equation}
The M-theory membrane has for its bosonic part the action
\begin{equation}\label{}
S_2=-T_2^M\int_{W_3}d^{3}\xi \sqrt{\det  
G^{M}_{ij}}-T_2^M\int_{W_3}C_{3},
\end{equation}
where the appropriate pull-back maps to the world volume are  
understood.
Let us first consider the case of double dimensional reduction where  
the
membrane is wrapped around the eleventh dimension to give a string.  
Using the
pull-back of the Kaluza-Klein ansatze (and choosing $W_3$ coordinates  
$\xi^i =
(\s^1,\s^2,\rho )$ with winding number $\pa_{\rho}y=\nu\e{\cal Z}
{}~(\pa_{\rho}X^{\mu}=0)$, one gets \cite{ptrev}
the string action

\begin{equation}\label{}
S_1=-\nu 2\pi\sqrt{\a '}T_2^M\int_{W_2}d^{3}\xi \sqrt{\det  
G_{ij}}-\nu
2\pi\sqrt{\a '}T_2^M\int_{W_2}B_{2}.
\end{equation}
Since this must be the  fundamantal string with tension (charge)  
${1/2\pi\a '}$
we get
\begin{equation}\label{memtension}
T_2^M={1\over (2\pi)^2\a '^{3\over 2}}.
\end{equation}
On the other hand by simple dimensional reduction one can get the  
D-membrane in
ten dimensional IIA string theory \cite{ptrev}. The gauge field on  
the D-brane
is actually the dual of the one form $dy$ coming from the  
compactified eleventh
coordinate field on the membrane \cite{dl}, \cite{pt},\cite{cs}.  
Specifically,
a term
\begin{equation}\label{}
-T_2^M(2\pi\a ')\int_{W_3} F_2\w (Y_1-C_1)
\end{equation}
where locally $F_2=dA_1$, is added. Integrating $A_1$ gives the  
condition
$d(Y-C)=0$ which is solved by $Y-C=dy$. This gives the original form  
of the
membrane action on $M_{10}\times S_1$ with the background metric. The
normalization is the correct one because  the integral cohomolgy  
classes are
defined as

\begin{equation}\label{}
{[F]\over 2\pi}\e{\cal Z},~~~~{[Y-C]\over 2\pi\sqrt{\a '}}\e{\cal Z}
\end{equation}
The first is the standard normalization of the gauge field and the  
second is
our standard identification of the eleventh coordinate  
(\ref{yidentify}). Then
using (\ref{memtension}) we see
that the additional term is just $2\pi$ times an integer. By  
integrating out
the field $Y$ one obtains

\begin{equation}\label{2brane}
S_2=-T_2^M\int_{W_3}d^{3}\xi e^{-\phi} \sqrt{\det (G_{ij}+{{\cal
F}_2}_{ij})}-T_2^M\int_{W_3}(A_{3}+{{\cal F}_2}\w C_1),
\end{equation}
where ${{\cal F}_2}\equiv 2\pi\a 'F_2-B_2.$
This is exactly the D-membrane action\footnote{This agreement  
\cite{cs} is only
shown at the classical level, and it does not seem possible to extend  
for
instance the argument made for mapping the type IIB SL(2,Z) dual  
strings to
each other quantum mechanically, to the present case \cite{sato}. }  
so that we
have from (\ref{memtension})
\begin{equation}\label{memtension2}
T_2=T_2^M={1\over (2\pi)^2\a '^{3\over 2}}.
\end{equation}
The consistency of the above expression with the formula  
(\ref{ptension}) is a
reflection of the agreement  of our conventions for T-duality and the   
standard
identification of the eleventh coordinate (\ref{yidentify}). Hence we  
may use
our earlier calculation of the ten dimensional gravitational constant
(\ref{kappa}) and (\ref{mtwoa}) to get,
\begin{equation}\label{kappaeleven}
2\k_{11}^2=(2\pi )^8\a '^{9\over 2}.
\end{equation}
It is also easy to see, by identifying the M-theory five-brane  
wrapped around
the circular eleventh dimension, with the D-four-brane that,  
$2\pi\sqrt{\a
'}T_5^{M}=T_4=1/(2\pi)^4\a '^{5\over 2}$ so that  
$T_5^{M}=1/(2\pi)^4\a
'^3=T_5$. This provides another check on the above relation for   
$\k_{11}$
since

\begin{equation}\label{}
2\k_{11}^2T^M_2T^M_5=(2\pi )^8\a '^{9\over 2}{1\over (2\pi)^2\a  
'^{3\over
2}}{1\over (2\pi)^2\a '^{3\over 2}}=2\pi,
\end{equation}
which is exactly the M-theory Dirac quantization condition.
For completeness we note finally that the Kaluza-Klein particle mass  
quantum
which in M-theory units is $\tau_0^{(M)} =1/R_{11}$ in string units  
has  mass
$1/(g_A\sqrt{\a '})=\tau_0$ \cite{ew1}, the mass of a D-0-brane.

Let us now check the consistency of the above with the normalizations  
of the
topological terms in  string theory effective actions. In the case of  
IIA the
latter is a result of the proper normalization of the topological  
term in the
M-theory action which was checked in \cite{ew}. The term in question  
is
(see (\ref{twoa}))
\begin{equation}\label{}
{1\over 2\k^2}\shalf\int_{M_{10}}F'_4\w F'_4\w B_2.
\end{equation}
By standard arguments the quantum consistency of the IIA low energy
action\footnote{It might be asked why one demands this, since the  
action in
question is already an effective action. The point is that this is an  
effective
action only to the extent that substring scale fluctuations have been
integrated out. However to discuss physics well below the string  
scale,
(presumably after compactification), one needs to integrate down  
further and
one therefore should require the above action to be quantum  
mechanically
consistent. } requires that the integral ${1\over  
2\k^2}\shalf\int_{M}F'_4\w
F'_4\w H_3$ (where $M$ is a closed eleven dimensional
manifold\footnote{Strictly speaking we should be talking about  
Euclidean
manifolds and the term should appear with a factor $i$.}, be $2\pi$  
times an
integer. From (\ref{magnetic}) we have the magnetic charge equations,
\begin{equation}\label{}
\int_{S_4}F'_4={2\pi\over T_2},~~\int_{S_3}H_3={2\pi\over T_1}.
\end{equation}
Choosing $M=S_4\times S_4\times S_3$ we have

\begin{equation}\label{}
{1\over 2\k^2}\shalf\int_{M}F'_4\w F'_4\w H_3={1\over 2\k^2}\shalf  
2\left
({2\pi\over T_2}\right )^2{2\pi\over T_1}={[(2\pi )(2\pi)^2\a  
'^{3/2}]^2(2\pi
)(2\pi \a')\over (2\pi )^7\a '^4 }=2\pi.
\end{equation}
In the first equality above the factor two comes from the two ways of
integrating $F_4$ over the
two $S_4$'s,  the next equality is obtained by substituting from  
(\ref{kappa})
and (\ref{ptension}), and the last equality shows the consistency of  
the
tension formulae with the normalization of the topological  
terms.\footnote{It
should be stressed that this is not a check of the question of  
whether the
toplogical term is well-defined. For that one needs to consider a  
general
tweleve manifold, and this is done in \cite{ew}.}

There is a similar topological term in the type IIB effective action.  
In this
case of course there is no direct connection to the M-theory action,  
but there
is an alternate argument\footnote{See \cite{ptrev} and references  
therein.}
which we summarize here. Type IIB has two two-form fields $B_2$ and  
$C_2$ from
the NS-NS  and R-R sectors respectively, with field strengths  
$H_3=dB_2$ and
$F_3=dC_2$. In addition this theory has a four form field $C_4^+$  
whose field
strength $F_5=*F_5$, i.e. is self--dual. The Bianchi identity  
satisfied by this
field is $dF_5=H_3\w F_3$ so that $F_5 = F'_5 -H_3\w C_2$, where  
locally we may
write $F'_5=dC_4$.\footnote{ See next section for a discussion of the  
gauge
invariance of such terms.} Now if one imposes the self duality on the  
fields in
the Lagrangian it is not possible to write a kinetic term for $C_4$.  
Instead
one imposes the self duality constraint only at the level of the  
equations of
motion (i.e. only the solutions must satisfy this constraint); but  
one then
needs to  add a topolgical term to the action in order that the  
equations of
motion be consistent with the Bianchi identity.
The terms in question are
\begin{equation}\label{}
{1\over 2\k^2}\int_{M_{10}}\left (\shalf F_5\w *F_5 +C_4\w H_3\w F_3  
\right ).
\end{equation}
The equation of motion $d*F'_5=H_3\w F_3$ is then the same as the  
Bianchi
identity once the
self-duality constraint is imposed. Now we have the magnetic charge  
equations,

\begin{equation}\label{}
\int_{S_3}H_3 = {2\pi\over T_1},~~\int_{S_3}F_3 = {2\pi\over
T_1},~~\int_{S_5}F_5={2\pi\over T_3}.
\end{equation}
Then as in the previous argument for the type IIA case we have
\begin{equation}\label{normtwob}
{1\over 2\k^2}\int_{S_5\times S_3\times S_3}F_5\w H_3\w F_3={1\over
2\k^2}{2\pi\over T_3}\left ({2\pi\over T_1}\right )^2 =2\pi,
\end{equation}
where we again used (\ref{ptension}), (\ref{kappa}) to get the last  
equality.
Note that on the left hand side of the first equality there is no  
factor two
unlike in the IIA case. This is because now we have two different  
three forms
in the integrand. Thus for instance if we
choose $y^1,y^2,y^3$ as the coordinates on one $S_3$ and $y^4,y^5,y^6  
$ on the
other, a minimal configuration is obtained by taking   
$H_3=H_{123}dy^1\w dy^2\w
dy^3,~F_3=F_{456}dy^4\w dy^5\w dy^6 $ whereas in the previous (IIA)  
case a
minimal configuration for the three form field
strength is  $F_3=F_{123}dy^1\w dy^2\w dy^3+F_{456}dy^4\w dy^5\w dy^6  
$.  The
last equality in (\ref{normtwob}) comes from substituting for $\k$  
and the
tensions from (\ref{kappa}) and (\ref{ptension}) and checks the  
consistency of
the normalization.

\sect{Nonrenormalization of  D-brane actions}

The bosonic part of the D-p-brane action is given by,
\begin{equation}\label{Dpbrane}
S_p=-T_p\int_{W_{p+1}}d^{3}\xi e^{-\phi} \sqrt{\det (G_{ij}+{{\cal
F}}_{ij})}-T_p\int_{W_{p+1}}C\w e^{{\cal F}_2}
\end{equation}
In the above $C$ stands for a formal sum of anti-symmetric tensors  
and ${\cal
F}_2$ is given in the line after (\ref{2brane}) and the appropriate  
sum of
wedge products of  $C_r$ and ${\cal F}_2$ coming from the expansion  
of the
exponential is understood . This form of the action can actually be  
derived in
pertubation theory from the disc topology \cite{clny},\cite{rl},
\cite{li},\cite{md}, \cite{cs}\cite{sato}. However we would like to
argue that it is valid as a  quantum exact but low energy and  
constant gauge
field strength, effective action coming from integrating out  
sub-string scale
fluctuations.

The action has a gauge invariance under, $C_r\rightarrow   
C_r+d\L_{r-1}-H_3\w
\L_{r-3} $.
As is well known a topological term is not expected to get  
renormalized since
for instance any  $\phi$ dependence will spoil the gauge invariance  
of this
term,\footnote{Even in field theory with a constant coupling such  
terms are not
renormalized since the normalization is fixed by topological  
considerations.}
so one does not expect the second term in the action to be  
renormalized. As for
the first term one needs to consider the following facts.  As argued  
in section
one the dilaton dependence when ${\cal F}=0$ can be obtained by  
imposing some
naturalness criteria and requiring the existence of solutions to the  
coupled
equations of background plus brane.  The question then is whether the
Born-Infeld piece $\sqrt{\det (1+G^{-1}{\cal F}}$ is valid (upto  
derivative
terms in the field strength) without acquiring (dilaton dependent)
renormalization.

The Born-Infeld (D-9-brane) action in ten dimensions  was originally  
derived
from  (open) string theory by calculating the partition function on  
the disc by
Fradkin and Tseytlin \cite{ft}. Subsequently however it was shown by  
Aboulsaoud
et al. \cite{acny} that the result can be obtained from the  
requirement of
conformal invariance i.e. from a beta function calculation.  
Consequently, (as
observed in that paper), one can argue that the result is independent  
of world
sheet topology i.e. it should be valid to all orders in perturbation  
theory.
The beta function argument was extended to $p<9$ branes by Leigh  
\cite{rl}. The
upshot is that
at least to all orders in perturbation theory the form of the DBI  
action should
be valid for all
D-branes.

What about non-perturbative renormalization? For  the  D-2-brane one  
might
think that one can go further since  the D-brane action can be  
obtained from
the M-theory 2-brane action. The latter  is the strong coupling  
version of the
former so  that this fact might be
interpreted as evidence for the non-renormalization (even  
non-perturbatively)
of the former. However the equivalence is shown only at the level of  
treating
some of the fields classically (see footnote 10) so that it is not  
clear that
this establishes the needed result.
 However recent work has given some support to the conjecture that  
the form of
the D-brane action is unrenormalized even non-perturbatively. This is    
the
demonstration \cite{s},\cite{bt} that the D-brane action must have a
$\k$-symmetry which ensures that it
has the right  number  of degrees of freedom consistent with  
supersymmetry.
Given the above
argument for the non-renormalization of the toplological term,  the
kappa symmetry seems to ensure that the Born-Infeld factor is also
unrenormalized.

When N D-branes come together  one expects the $U(1)^N$ symmetry to  
be enhanced
to
$U(N)$ \cite{ew2}. Since in the limit when the gauge field  
configuration is in
the Cartan sub algebra we should get for the sum of the diagonal  
terms each of
which is the abelian action the  D-brane action should get replaced  
by
\begin{equation}\label{Dpbrane}
S_p=-T_p\int_{W_{p+1}}d^{3}\xi e^{-\phi} S\tr\sqrt{\det  
(G_{ij}+{{\cal
F}_2}_{ij})}-T_p\int_{W_{p+1}}C\w \tr e^{{\cal F}_2}+\ldots
\end{equation}
where the error  would be terms which vanish in the abelian limit.  
i.e.
commutator terms (including terms for the dynamical transverse  
position
variables of the D-brane) . (See \cite{at} for a detailed  
discussion). Thus up
to such terms the above should be a valid quantum
effective action for N concident D-branes.  In the next section  we  
will use
these results (for $p=9$)   to argue
that gauge coupling terms in type one and heterotic theories are  
fixed
non-perturbatively at their lowest order values.

 \sect{\bf Type I strings, heterotic strings and M-theory}
So far our discussion has been limited to type IIA, M-theory on  
smooth
manifolds, and type
IIB string theory. Let us now extend it to type I and heterotic  
strings. The
discussion is a modified and extended version of that given in  
\cite{sda1}.

Type I strings are obtained from type IIB by making an orientation  
projection
and adding (for consistency) SO(32) open strings D-ninebranes (see  
\cite{jprev}
and references therein).
To proceed we need to assume that the relation between the string  
scale $\sqrt
{\a '}$ and  $\k$ is the same as in type II. Since the projection  
leading to
type I  from type II does not affect the gravitational interactions  
we believe
this is a very plausible assumption.   The low energy effective  
action for this
theory may  be written,

\begin{eqnarray}\label{typeI}
I_I&=&-{1\over (2\pi )^7\a  
'^4}\int_{M_{10}}[e^{-2\phi}\sqrt{-G}(R-4(\nabla\phi
)^2)-\shalf F_3\w *F_3]\nn
& &-{1\over (2\pi)^9\a '^5}\int_{M_{10}}e^{-\phi}\tr (\sqrt{\det
(G_{\mu\nu}-2\pi\a 'F_{\mu\nu})}-\sqrt{ -G})\nn & & -{(2\pi\a  
')^4\over
(2\pi)^9\a '^5}\int_{M_{10}}C_2\w \tr F_2^4+\ldots
\end{eqnarray}
In the last two lines we have the 9-D-brane action with its tension  
given by
(\ref{ptension}).

The gauge field terms come from the D-9-brane and are not expected to  
get
renormalized as argued in section 4.
Note that the leading (two derivative) gauge field term is ,

\begin{equation}\label{}
-{1\over 4(2\pi)^7\a '^3}\int_{M_{10}}\sqrt{ -G}\tr F^2
\end{equation}

By the field redefinitions, $G_{\mu\nu}=e^{-\phi_H}G_{H\mu\nu}$ and
$e^{\phi}=e^{-\phi_H}$
 we get the low energy effective action of the SO(32) heterotic  
string theory:

\begin{eqnarray}\label{H32}
I_{H}&=&-{1\over (2\pi )^7\a  
'^4}\int_{M_{10}}\sqrt{-G}e^{-2\phi_H}\left\{
R-4(\nabla\phi )^2+{1\over12}F_3^2 \right\}\nn
& &-{1\over (2\pi)^9\a '^5}\int_{M_{10}}e^{\phi_H}\tr (\sqrt{\det
(e^{-\phi_H}G_{H\mu\nu}-2\pi\a 'F_{\mu\nu})}-e^{-5\phi_H}\sqrt{  
-G_H})\nn & &
-{(2\pi\a ')^4\over (2\pi)^9\a '^5}\int_{M_{10}}C_2\w \tr F_2^4
\end{eqnarray}
(In the above the $R$ and the contractions are defined with respect  
to $G_H$).
Now the leading gauge field term is
\begin{equation}\label{}
-{1\over 4(2\pi)^7\a '^3}\int_{M_{10}}\sqrt{ -G_H}e^{-2\phi_H}\tr F^2
\end{equation}
 Note that the ratio of the gauge field coupling to the gravitational  
coupling
is exactly as predicted to all orders in perturbation theory.
It is instructive to expand the square root determinant terms in  
these two
actions to $O(F^4)$ and then compare the results with standard string
perturbation theory calculations. To the extent that the terms have  
been
calculated they are a check on S-duality \cite{at'}.

Now the $E_8\times E_8$ heterotic theory is related to the $SO(32)$  
one by
T-duality. Hence the leading terms in this action will be

\begin{eqnarray}\label{HE8}
I_{H'}&=&-{1\over (2\pi )^7\a
'^4}\int_{M_{10}}\sqrt{-G_{H'}}e^{-2\phi_{H'}}\left\{ R-4(\nabla\phi
)^2+{1\over12}F_3^2 \right\}\nn
& &-\sum_i{1\over 4(2\pi)^7\a '^3}\int_{M_{10}}\sqrt{
-G_{H'}}e^{-2\phi_{H'}}\tr F_i^2
\end{eqnarray}
Note that as far as the gravitational terms go this is the same  
transformation
as that taking us from type IIB to type IIA.
 One takes a configuration with an isometry and compactifies in that  
direction
on a circle of radius R. The T-dual theory is then on a circle of  
radius
$R'={\a'\over R}$ and the dilaton is related by $e^{\phi_{H'}}={\sqrt  
{\a
'}\over R}e^{\phi_{H}}$. In the present case however one has also to  
introduce
a Wilson line which breaks the $SO(32)$ symmetry to $SO(16)\times  
SO(16)$. Upon
letting $R'\rightarrow\infty$ one gets a  ten-dimensional theory with
$E_8\times E_8$ symmetry\footnote{We note that the further  
transformation
$G_{H'}=e^{\phi_{H'}}G'_{\mu\nu},~e^{\phi '}=e^{\phi_{H'}}$ gives us  
a form of
the action that is similar to  type I and maybe  called type  
$\overline I$.
\begin{eqnarray}
I_{\overline I}&=&-{1\over (2\pi )^7\a
'^4}\int_{M_{10}}[e^{-2\phi}\sqrt{-G'}(R-4(\nabla\phi )^2)-\shalf  
F_3\w
*F_3]\nn
& &-\sum_i{1\over 4(2\pi )^7\a '^3}\int_{M^i_{10}}\sqrt{ -G'}e^{-\phi  
'}\tr
F_i^2
  \end{eqnarray}
One  might  have expected that  the above is the effective low energy   
action
coming from the type IA (or I') theory discussed in \cite{pw}.  
However this
does not appear to be the case. The type IA theory with coupling  
constant
$e^{\phi_{IA}}$ compactified on a circle of radius $R_{IA}$  is  
related to the
$E_8$ theory with coupling constant $e^{\phi_{H'}}$ on a circle of  
radius $R_8$
by the relations $R^2_{IA}=e^{\phi_{H'}}R_{8},~e^{{2\over  
3}\phi_{IA}}=
R_8 e^{-{1\over 3}\phi_{H'}}$ which is not in agreement with the  
above
relations to what we have
called ${\overline I}$.} \cite{nsw},\cite{pg1}. The gauge to  
gravitational
coupling ratio is fixed to all orders in  heterotic perturbation  
theory by a
well-known argument \cite{pg2}.

The strong coupling version of this theory has been identified  
\cite{hw} as M
theory on an orbifold, $M_{10}\times {S_1\over Z_2}$ by Horava and  
Witten.
These authors  determined (by requiring gauge anomaly cancellation)  
the
dimensionless ratio of gauge to gravitational constants,(see also  
\cite{sda2}).
With our value for $\k_{11}$ the action in question (in upstairs form  
- i.e. on
the smooth manifold $M_{10}\times S_1$ with a $Z_2$ symmetry on the  
fields ) is
(omitting the topological terms)
\begin{eqnarray}\label{HWM}
S_{H'}&=&-{1\over (2\pi )^8\a '^{9\over 2}}\int_{M}\sqrt{-G_M}\left\{  
R+{1\over
48}K_4^2 \right\}\nn
& &-\sum_i{1\over 4(2\pi )^7\a '^3}\int_{M^i_{10}}\sqrt{
-G_{H'}}e^{-2\phi_{H'}}\tr F_i^2
\end{eqnarray}
 In the above $M^i_{10}, ~i=1,2$ are the ten dimensional manifolds at  
the $Z_2$
fixed points on which the gauge fields live.

Let us now check the calculation of the gauge coupling constant by  
comparing to
the standard form of the heterotic effective action. The metric  
transformation
that needs to be done is almost the
same as that which takes us from the M-theory to IIA, namely,
\begin{equation}\label{}
ds^2_M=e^{-{2\over 3}\phi_{H'}  
(x)}G_{H'\mu\nu}(x)dx^{\mu}dx^{\nu}+e^{{4\over
3}\phi_{H'} (x)}dy^2
\end{equation}
 (On the fixed ten dimensional planes the RR fields $C_1, C_3$  
restricted to
these planes  disappear because of the symmetry (under for instance
$y(=x^{11})\rightarrow -y)$). Then (under the assumption that the  
fields are
independent of the eleventh coordinate) the action (\ref{HWM}) goes  
over into
(\ref{HE8}). Note that the fact that the gauge coupling that was  
fixed at the
M-theory level by anomaly cancellation, goes over exactly to the  
gauge coupling
as fixed in string theory, is an independent check on the  
Horava-Witten theory.

\sect{2-brane and 5-brane couplings in M-theory }

In this section we discuss the couplings and normalizations of the  
membrane and
the five-brane
of M-theory. In particular we discuss how to obtain an action
 which incorporates five branes and membranes  (with the latter  
having
possible boundaries sitting on five-branes \cite{as}\cite{pt}),   
coupled to
11-D supergravity. This demonstrates the consistency of the picture  
developed
for
these interactions
and also suggests a possible resolution to a problem with five-brane  
anomalies
pointed out recently \cite{ew4}. The discussion will in effect extend  
earlier
work
\cite{pt},\cite{ew3},\cite{oa},\cite{bro},\cite{ew4}, on these  
matters.

The low energy effective action of M-theory is given in its
standard form by equation (\ref{mnormal}).
It is to this  form of the supergravity action that the membrane  
couples
naturally.  There should however be another dual form of the action  
in terms of
a seven form field strength $K_7=*K_4 = dC_6+\ldots$  to which the  
five-brane
which is the magnetic dual of the membrane  naturally couples.  
Because of the
existence of the gauge field dependent topological term in  
(\ref{mnormal}) the
standard dualization procedure cannot be carried out, and the  
complete low
energy action for M-theory coupling to five branes has not been
given.\footnote{Another reason is the existence of a self-dual field  
strength
on the five-brane. }  We hope in this section to remedy this  by  
giving the
bosonic terms in this action. In fact we will   present one action  
from which
all the relevent equations can be obtained.

It is convenient to begin with the brane actions.
For a closed membrane there is in addtion to the volume term
$\int_{W_2}\sqrt{-det G_{ij}}$ (where $G_{ij}$ is the pull-back of  
the
M-theory metric to the world volume $W_2$ of the 2-brane) the     
coupling
$\int_{W_3} C_3$ to the three form field\footnote{Here and in all  
subsequent
formulae we will not explicitly write the pull-back map which is  
necessarily
there in order to write
the integral over the world volume of a field which is defined on the  
ambient
M-theory space.}
When the membrane has a boundary however this has to be modified   
since the
above term is no longer gauge invariant under the standard M-theory  
gauge
transformation $C_3\rightarrow C_3+d\L_2$. One needs then an  
additional field
on the boundary whose gauge transormation will cancel the boundary  
contribution
from  the above piece. Specifically we need

\begin{equation}
\int_{W_3} C_3 -\int_{\pa W_3}b_2
\end{equation}
 where $b_2$ is a two form field living on the two dimensional  
boundary of the
membrane world volume with the gauge transformation $b_2\rightarrow  
b_2+\L_2$ .
Where can this field come from? The only possibility is that the  
boundary sits
on a five-brane world volume \cite{pt},\cite{as}. The latter is known  
to have a
self-dual gauge field strength $h_3$. However we cannot just have
$h_3=db_2$ since this would not be invariant under the above gauge
transformation. The correct modification is clearly  \cite{pt},  
\cite{ptrev},
\begin{equation}\label{}
h_3= db_2-C_3|_{W_6}
\end{equation}
Note that this implies the Bianchi identity
\begin{equation}\label{}
dh_3=-dC_3|_{W_6} = -K_4|_{W_6}
\end{equation}
This also means, since $h_3$ must be globally defined on $W_6$, that
$T_2K_4/2\pi$ (which in general has integral or half integral  
periods) actually
has vanishing periods when restricted to
the world volume of the five-brane \cite{ew4}.

What then is the action for this field?
The problem  is that it is self-dual. We will adopt here the
strategy used in earlier work (see \cite{ptrev} and references  
therein), namely
to impose the self-duality constraint at the level of the equations  
of motion
and then require consistency with the Bianchi identity. Including  
also the
coupling to the six form field $C_6$ this turns out to be
\begin{equation}\label{5action}
-\shalf\int_{W_6}h_3\wedge *h_3 - 2 \int_{W_6}C_6+\int_{W_6}C_3\wedge  
db_2.
\end{equation}
 We will  show below that self duality of $h_3$ and gauge invariance  
determine
the (relative) coefficients  as written above\footnote{For recent   
discussion
of  the 5-brane action see \cite{blnpst}, \cite{apps}, \cite{hsw}.}.

The action is to be used
to derive the field equations and the self duality constraint $  
h_3=*h_3$ is
imposed at the level of the equations of  motion. We get, by varying  
with
respect to $b_2$,
\begin{equation}\label{}
d*h_3=-d C_3
\end{equation}
Using the self duality condition this is exactly the Bianchi  
identity. Thus the
relative coefficient of
the first and last terms is fixed \cite{ptrev}. Now consider the  
Bianchi
identity for  the seven form field strength\footnote{We shall ignore  
for the
moment an additional piece coming from gravitational anomalies on the
five-brane world volume \cite{dlm}}, $K_7$ (which is dual to $K_4$),  
which
follows from the equation of motion for $C_3$ (see (\ref{mnormal}).
\begin{equation}\label{}
dK_7=d*K_4=-\shalf K_4\wedge K_4
\end{equation}
The solution of this equation is
\begin{equation}\label{kseven}
K_7=dC_6-\shalf C_3\wedge K_4
\end{equation}
The invariance of the field strength $K_7$ under $\d C_3=d\L_2$ then  
gives $\d
C_6=\shalf \L_2\wedge K_4$. It is then easily checked that the gauge  
invariant
combination of topological terms is the one given in equation
(\ref{5action})\footnote{This seems to have been first pointed out in
\cite{oa}}. Restoring the five-brane tension we then have the  
complete bosonic
five-brane action

\begin{equation}\label{fiveaction}
S_5=T_5\int_{W_6}\sqrt{-det G_{ij}}-{T_5\over 4}\int_{W_6}h_3\wedge  
*h_3 - T_5
\int_{W_6}C_6+{T_5\over 2}\int_{W_6}C_3\wedge db_2.
\end{equation}

Now this five-brane cannot couple to the original form of the action  
where the
only independent
variable (apart from gravity) is the three form field. This would  
leave $C_6$
without a kinetic term and its equation of motion will imply that  
$T_5=0$. We
need the dual formulation of the 11 dimensional action in which
$K_7=dC_6+\ldots$ appears. The difficulty with writing this down is  
the
presence of the topological term  which involves $C_3$ explicitly and  
so to our
knowledge such an action has not been written down before. The  
following is
perhaps a step toward solving this problem.
 Treating $C_3, C_6$ as independent fields we write down the  
(bosonic)  action,
(omitting the pure gravity piece).

\begin{equation}\label{mdual}
\tilde I_{11}=-\shalf\int K_7\w *K_7+{1\over 3}\int C_3\w dC_3\w dC_3
\end{equation}
The integrals in the above are taken over $M$ as in
(\ref{mnormal}).\footnote{In the rest of this section an integral  
whose range
is not indicated is to be taken over $M$.} Note that the coefficient  
of the
topological term is twice that of (\ref{mnormal}). $K_7$ in the above  
action is
defined by equation (\ref{kseven}) with $K_4=dC_3$. Now this action  
involves
both $C_3$ and $C_6$ so it is not what one would normally call a dual  
action.
In fact it is probably impossible to construct an action involving  
$C_6$ alone
since the Bianchi identity for $K_7$ involves $C_3$. Nevertheless it  
is on a
similar  footing to the action for IIB supergravity\cite{ptrev}  
(which involves
a self dual five form field strength) or the action  
(\ref{fiveaction}) which
involves a self-dual three form. As in those cases we need to impose  
a
constraint, at the level of the field equations, that is consistent  
with them.
In the present case this constraint is $*K_7=-dC_3$.   The variation   
of the
action with respect
to
$C_6$  gives the equation of motion  $d*K_7=0$. Variation with  
respect to $C_3$
gives, after using
the previous equation, $dC_3\w (*K_7+dC_3)=0$. This is clearly  
consistent with
our constraint. \footnote{I wish to thank  Eugen Cremmer, Bernard  
Julia, Hong
Lu, and Chris Pope for pointing  out an incorrect statement in a  
previous
version of this argument.}

The above action enables us to consider $C_6$ as a dynamical field  
and now we
may couple two and five branes. (It is not obvious how to do this   
with the
original
form of the
M-theory action (\ref{mnormal})). Thus we have the action $I= \tilde  
I_{11}
+S_2+S_5$ where the first term on the right hand side is given by
(\ref{mdual}), the second by $-T_2\int_{W_3}C_3+T_2\int_{\pa W_3}  
b_2$, and the
third by (\ref{fiveaction}). The independent fields are $C_3, C_6,$  
and $b_2$.

It turns out that the definition of $h_3$ needs to be modified  
because of the
non-zero boundary of $W_3$ i.e. we now must have
$h_3=db_2-C_3|_{W_6}+{2T_2\over T_5}\t_3 (W_6\rightarrow W_3)$  where  
$\t_3$
the restriction of $M$ to $W_3$ is defined below.

In order to derive the equations of motion in the presence of these  
sources it
is convenient to
introduce delta (step) function differential forms which enable us to  
write all
the terms as
integrals over $M$. Thus we put for the integral of an arbitrary  
6-form $A_6$,
(for which we assume that an extension to $M$ with compact support  
exists),
\begin{equation}\label{extension}
\int_{W_6}A_6 =\int_{M}A_6\w \d_{5}
\end{equation}
Here $\d_{5}$ is a closed delta function 5-form which is a natural
genaralization of the delta one-form $\d (x)dx$ (see (\ref{delta}),
(\ref{exdelta})). Note that this makes sense only if $W_6$ is
closed\footnote{Consider $A_6=d\L_5$. Then since $\d$ is closed the  
right hand
side vanishes for arbitrary $\L_5$ vanishing at infinity in $M$, but  
the left
hand side vanishes only if $W_6$ is closed.} (i.e. $\pa W_6  
=\emptyset$. By
contrast consider the seven manifold $W_7^+$ which has a non-zero  
boundary
$W_6$. Now the restriction must be  given by a generalization of a  
stepfunction
times a  delta function form $\t_4$ which has compact support in the  
directions
normal to $W_7^+$ .  Let us  integrate a seven form $F_7=dA_6$ over  
$W_7^+$
 \begin{equation}\label{}
\int_{W_7^+}dA_6=\int_{M}dA_6\w \t_4 =-\int_{M}A_6\w  d\t_4.
\end{equation}
But from Stokes' theorem the left hand side is equal to the left hand  
side of
(\ref{extension}). Hence  consistency  requires  $d\t_4=-\d_5$. We  
may deal
with the integral
over $W_3$ (whose boundary is non-null)  in a similar fashion i.e. by  
writing
$\int_{W_3}C_3=\int_M C_3\w\t_8$.

 Now the  equations of motion for the combined action may be derived  
and we
get,
\begin{eqnarray}\label{}
\d C_6: ~~d*K_7&=&T_5\d_5(M\rightarrow W_6)\nn
\d C_3:~~ -dC_3\w*K_7&=&dC_3\w dC_3-2T_2\t_8({M\rightarrow
W_3})+T_5h_3\w\d_5\nn
\d b_2:~~dh_3&=&-dC_3|_{W_6}-{2T_2\over T_5}\d_4 (W_6\rightarrow\pa  
W_3).
\end{eqnarray}
In the above we have again used the self-duality constraint for  
$h_3$. The
constraint relating $C_6$ and $C_3$ has to be changed now to be  
consistent with
the second equation. Thus we need to put  $*K_7=-dC_3-T_5\t_4$ where
$d\t_4=-\d_5(M\rightarrow W_6)$. Similarly the
modification of the definintion of $h_3$ mentioned earlier is  
necessary in
order to have compatibility with the last equation. Note that taking  
the
exterior derivative of the second equation and using  the first,   
yields an
equation that is consistent with the third.

Let us now write down a self-dual form of the M-theory action.
\begin{equation}\label{sdeleven}
I_{sd}=-\shalf\int K_4\w *K_4-{1\over 6}\int C_3\w dC_3\w dC_3 +\int  
K_7\w
(K_4-dC_3).
\end{equation}
Here $K_4,K_7$ and $C_3$ are to be treated as independent fields.
Thus we have the following field equations.

\begin{eqnarray}\label{sdeqns}
\d_{K_7}:~~~K_4&=&dC_3,\nn
\d_{K_4}:~~~*K_4&=&K_7,\nn
\d_{C_3}:~~~dK_7&=&-\shalf dC_3\w dC_3.
\end{eqnarray}
If the first equation is substituted into the action (\ref{sdeleven})  
we get
the normal form of the action i.e. equation (\ref{mnormal}). If on  
the other
hand   in (\ref{sdeleven})  we use the second equation (and its dual
$K_4=-*K_7)$ as well as the solution $K_7=dC_6-\shalf C_3\w dC_3$ to  
the
constraint given in the last equation,  we get the dual form  
(\ref{mdual}).

Now consider the coupling of 2- and 5-branes to this action.  The  
action
becomes,

\begin{eqnarray}\label{msdbrane}
I_{sd}+S_2+S_5 &=&-\shalf\int_M K_4\w *K_4-{1\over 6}\int_M C_3\w  
dC_3\w dC_3
+\int_M K_7\w (K_4-dC_3)\nn & &-T_2\int_{W_3}C_3+T_2\int_{\pa W_3}  
b_2
-{T_5\over 4}\int_{W_6}h_3\w *h_3\nn& &-T_5\int_{W_7^+}(K_7+\shalf  
K_4\w
C_3)+{T_5\over 2}\int_{W_7^+}K_4\w db_2
\end{eqnarray}
In the above we take $M$ and $W_6$ to be closed manifolds but $\pa  
W_3\ne 0$.
The manifold $W_7^+$ is such that its boundary $\pa W_7^+=W_6$. Note  
that the
integrals over
$W_7^+$ are in fact a rewriting of the topological integrals over  
$W_6$ in
(\ref{fiveaction}). In the above action however the independent  
fields are
taken to be $C_3,K_4,K_7 $ and $b_2$. Note that this form of the  
action (i.e.
involving $K_7$ rather than $C_6$ (and hence also an integral over  
the open
disc $W_7$ rather than the closed manifold $W_6$) is forced upon us  
by the
requirement of a consistent coupling to the M-theory background. As  
before we
define $h_3=db_2-C_3+(2T_2/T_5)\t_3$. The action is invariant under    
the gauge
transformations $C_3\rightarrow C_3+d\L_2,~b_2\rightarrow b_2+ \L_2,
h_3\rightarrow h_3, K_4\rightarrow K_4, K_7\rightarrow K_7. $

Now   the field equations are,
\begin{eqnarray}\label{eom}
\d_{K_7}:~~~K_4&=&dC_3+T_5\t_4\nn
\d_{K_4}:~~~*K_4&=&K_7+{T_5\over 2}h_3\w\t_4\nn
\d_{C_3:}~~~dK_7&=&-\shalf dC_3\w dC_3-T_2\t_8+{T_5\over 2}(h_3\w
\d_5-K_4\w\t_4)\nn
\d_{b_2}:~~~dh_3&=&-K_4|_{W_6}-{2T_2\over T_5}\d_4(W_6\rightarrow\pa  
W_3)
\end{eqnarray}
In the last two equations we have used the self-duality constraint  
$h_3=*h_3$.
Note that the consistency condition for the third equation,  
$d^2K_7=0$, follows
from the fourth equation and the use of $\t_4(M\rightarrow
W^+_7)\w\d_5(M\rightarrow W_6)=0$.

We have demonstrated above  the consistency of the picture that has  
been
diccussed in the
literature on the coupling of two- and five-branes in M-theory.  The  
self-dual
form in fact allows one to reexpress these couplings in terms of the  
original
$K_4$ form of the supergravity action. Thus by substituting the first  
of
equation (\ref{eom}) into the self-dual action (\ref {msdbrane}) we  
have
\begin{eqnarray}\label{kfour}
I&=&-\shalf\int_{M}K_4\w *K_4-{1\over 6}\int_M C_3\w dC_3\w dC_3\nn &
&-T_2\int_{W_3}C_3+T_2\int_{\pa W_3} b_2 -{T_5\over 4}\int_{W_6}h_3\w  
*h_3\nn &
&-{T_5\over 2}\int_{W_7^+}dC_3\w h_3,
\end{eqnarray}
with $K_4=dC_3+T_5\t_4$.

Note that even though in the dual form the couplings to the  
five-brane can be
expressed in a
purely six-dimensional form in the above we have to use explicitly  
the seven
manifold $W_7^+$.
 This is the  analog of the Dirac string. We will show in the next  
section
that  this and similar terms coming from perturbative anomalies  can  
be written
in a form that is manifestly independent of the seven  
manifold.\footnote{I wish
to thank E. Witten for emphasizing the necessity of showing this.}

\sect{\bf Perturbative anomalies in the presence of five-branes}

The  self dual form of the M-theory low energy action and its  
coupling to
two and five branes, enables us now to discuss the origin  of  a  
certain term
that is necessary for the cancellation of  (gravitational) anomalies  
in the
Horava-Witten theory.
This  term is required because of the existence of gravitational  
(local
Lorentz) anomalies on the 6D world volume of the five brane.  The  
original
discussion of this term has to be modified in the light of the  
observations in
\cite{ew4} so let us brefly review those.

Let $TM$ be the tangent bundle on $M$. Its restriction to $W_6$ may  
be written
as $TW\oplus N$ where the first term is the tangent bundle  and the  
second the
normal bundle on $W_6$.
The anomalies in question are associated with  local Lorentz  
tranformations
preserving the direct sum structure of the bundle and come from two  
sources.
They are the chiral fermion on the five-brane which couples to the  
spin
connection on the brane world volume, and an $SO(5)$ gauge field  
associated
with  rotations of the normal bundle, and the chiral two form field  
$b$ which
`sees' only tangent bundle rotations. As usual the anomaly is  
associated with
index theory and characteristic classes on a manifold $W_8$ in two  
higher
dimensions. As computed in \cite{ew4} the chiral fermion anomaly is  
related by
the usual descent formalism to the following eight form,

\begin{equation}\label{index}
\shalf I_{Dirac}(TW\oplus N)=\shalf\left  
[4{7(p_1(TW)/2)^2-p_2(TW)\over
1440}-{p_1(N)p_1(TW)\over 48}+{p_1^2(N)\over 96} + {p_2(N)\over  
24}\right ]
\end{equation}

where $p_i(TW),~p_i(N),~~i=1,2$, are the first and second Pontryagin  
classes of
 the tangent and normal bundles on $W_6$. The integral of $I_{Dirac}$  
is the
Dirac index and the factor half is present because the fermion is  
(symplectic)
Majorana-Weyl.

The chiral two form  anomaly is related to,

\begin{equation}\label{}
I_A={1\over 5760}[16p_1^2(TW)-112p_2(TW)
\end{equation}

Define $\O_8={1\over 48}(p_2-(\l )^2)(TW\oplus N)$ (where  
$\l=p_1/2$).
Then using  $p_1(TM)= p_1(TW)+p_1(N),~p_2(TM) =  
p_2(TW)+p_1(TW)p_1(N)+p_2(N)$
one has the result  that
the total anomaly which needs to be cancelled  is descended from,
\begin{equation}\label{anomaly}
\shalf I_{Dirac}+I_A= -\O_8(TM)+{p_2(N)\over 24}
\end{equation}
As observed in \cite{ew4} the first term on the LHS can be cancelled  
by anomaly
inflow from
a term \cite{vw}\cite{dlm} proportional to  $\int K_4\w \O_7^{C.S.}$.
\footnote{This term is usually written as $\int C_3\w \O_8$ and in  
the absence
of five branes it is equal to the term written in the text. In the  
presence of
a five brane they are not the same because $K_4 \ne dC_3$.} Here
$\O_8=d\O_7^{C.S.}$, with $\d\O_7^{C.S.}=d\O_6^1$, so that using  
$dK_4=-T_5\d$
and Stokes' theorem we have the required anomaly cancelling  
variation. However
as pointed out
in that work the second term remains uncancelled.

A way to achieve anomaly cancellation  in the type IIA case was given  
in
\cite{ew4}. Let us first give a (slightly modified) account of this.  
When
$M=M_{10}\otimes S^1$ the normal bundle to the five-brane becomes
$N=N'\oplus O$ where $O$ is the trivial tangent bundle to $S^1$ so  
that
$p_i(N)=p_i(N')$. The structure group of $N'$ is $SO(4)$, so that  
$p_2(N')=\chi
(N')^2$, where $\chi$,  the Euler character of $N'$, is represented  
by $\chi
(F)=F^{ab}\w F^{cd}\e_{abcd}/32\pi^2$. The standard
descent equations then read, $\chi (F)=d\chi_3^{C.S.},~\d
\chi_3^{C.S.}=\chi_2^1$ where $\d$ is a local $SO(5)$ variation. The  
anomalous
variation that is to be cancelled is then proportional to  
$\chi_2^1\w\chi (F)$,
and hence can be cancelled by adding to the action the term
\begin{equation}\label{counter}
{2\pi\over 24}\int_{W_6}B|_{W}\w\chi (F),
\end{equation}
where the two form field $B$ of type IIA string theory, when  
restricted to
$W_6$ must now have the  variation $\d B=- \chi_2^1$ under local  
gauge
transformations. This however means (as in the usual Green-Schwarz  
mechanism)
that the gauge invariant field strength is redefined to be
$H|_{W_6}=dB|_{W_6}+\chi_3^{C.S.}$, which implies that  
$dH|_{W_6}=\chi (F)$.
The equation implies that a necessary condition for this mechanism to  
work is
that the  Euler character $\chi$, integrated over any four cycle on  
the
five-brane world volume, must vanish. In this case $\chi_3^{C.S.}$ is  
globally
defined and we may rewrite the counter term by using Stokes' theorem  
and the
definition of  $H|_{W_6}$ as
 \begin{equation}\label{countertwo}
{2\pi\over 24}\int_{W_6}H|_{W}\w\chi_3^{C.S.}
\end{equation}
which is the form given in \cite{ew4}. On the 10d manifold in the  
presence of
five branes,
the Bianchi identity for $H$ is modified to $dH=\d$ (where $\d$ is a  
delta
function four form with support on $W_6$) and it was observed in  
\cite{ew4}
that  when restricted to the five brane world volume, this   reduces  
to the
equation $dH|_{W_6}=\chi (F)$ given earlier,  as a consequence of the  
theory of
the Thom isomorphism.
As pointed out in \cite{ew4} it is not clear how to generalize this  
argument to
M-theory.   It was suggested there that one could use the formula
\begin{equation}\label{p2n}
p_2(N)=\sum_e\chi_e\w\chi_e
\end{equation}
where $\chi_e={F^{ab}\w F^{cd}\e_{abcde}/ 32\pi^2}$ with $a,\ldots e$  
being
$SO(5)$ indices in the normal bundle. It clearly reduces to the  
previous case
when  $N=N'\oplus O$.
In M-theory unlike in type IIA there is no three form field strength  
so it was
proposed  that one should use (on $W_6$)  the four form field $K$  
with one
index in the normal bundle. i.e.
introduce $K\sim H_a\w e^a$ where $H_a$ is a three form field and  
$e^a$ is a
vielbein one-form which is (vector) valued in the  normal bundle. It  
was
suggested in \cite{ew4} (in analogy with (\ref{countertwo})  that a  
counter
term of the form $\int_{W_6}$ could be introduced to cancel
the uncancelled anomaly in M-theory. Below we will try to work this  
suggestion
out in some detail.

First we note that the structure group of the tangent bundle to $M$  
in the
neighborhood of the five-brane is $SO(1,5)\otimes SO(5)$. Introducing  
a $SO(5)$
connection $A^{ab}$ on the normal bundle to the five-brane, we have  
the
covariant constancy equation for the vielbein one-form in the normal  
direction,
\begin{equation}\label{covcons}
de^a+A^{ab}e^b=0.
\end{equation}
 (Note that indices go over the vector representation of $SO(5)$ and  
are raised
and lowered using the  metric $\d_{ab}$ on the fiber).
The field strength $F^{ab} =dA^{ab}+A^{ac}\w A^{ca}$ is the curvature  
of the
normal bundle and satisfies the Bianchi identity  
$DF^{ab}=dA^{ab}+(A\w F-F\w
A)^{ab}$. Using this and the $SO(5)$ invariance of the epsilon symbol  
it
follows that $\chi_a$ is covariantly constant. i.e.
\begin{equation}\label{chicons}
(D\chi) _a = d\chi_a+A^{ab}\w\chi_b=0
\end{equation}
(From this it is easily checked that $dp_2(N)=0$ with $p_2$ given by
(\ref{p2n}), as should be the case of course).
Note that we have also have from the above the integrability  
conditions
\begin{equation}\label{consis}
F^{ab}\w e_b=F^{ab}\w \chi_b=0.
\end{equation}

Now we need to work out the relevant descent formalism. Consider the  
five form
$\chi\equiv \chi_a\w e^a$. From the covariant constancy of the two  
factors it
follows that $d\chi =0$. Hence there is a Chern-Simons type four  
form, which
may be written $\chi^{C.S.}_a\w e^a$ since it must scale linearly  
with  $e_a$,
such that
\begin{equation}\label{}
\chi_a\w e^a = d(\chi^{C.S.}_a\w e^a)  
=(d\chi^{C.S.}_a+A^{ab}\chi_b^{C.S.})\w
e^a
\end{equation}
where we have used (\ref{covcons}) and the the antisymmetry of the  
connection
matrix. Thus
we may set $\chi_a =d\chi^{C.S.}_a+A^{ab}\chi_b^{C.S.}$. Note that  
this works
because of (\ref{consis}) and that  it also implies
$F^{ab}\w \chi_b^{C.S.}=0$. Now since $\chi_a\w e^a$ is gauge  
invariant we
have,
\begin{equation}\label{}
\d (\chi^{C.S.}_a\w e^a)=d(\chi_a(\a, A)\w e^a)=(d\chi_a(\a,
A)+A^{ab}\chi_b(\a, A))\w e^a
\end{equation}
where we have again used the covariant constancy of $e^a$. In the  
above $\a$ is
a local $SO(5)$ gauge parameter so  $\d e^a =\a^{ab}e^b,~\d  
A^{ab}=[\a,
A]^{ab}+d\a^{ab}$. Thus we may write, $\d \chi^{C.S.}_a=d\chi_a(\a,
A)+A^{ab}\chi_a(\a, A) +(\a\chi(\a, A))_a$.
The anomaly coming from $p_2(N)$ via  this modified descent formalism  
is
\begin{equation}\label{des}
\sum_a \chi_a(\a, A)\w\chi_a(F).
\end{equation}

Now the component of $K$ on $W_6$ with one index in the normal bundle  
may be
witten as $H_a\w e^a$. Since this must be closed we have locally  
(using again
the covariant constancy of $e$) $H_a=dB_a+A_{ab}\w B_b$ where $B_a\w  
e^a\sim C$
is the component of the M-theory three form field $C$ with one index  
in the
normal bundle. The transformation $\d C=d\L$ induces the  
transformation $\d
B_a=(D\L )_a$ and hence $\d H_a =F_{ab}\w\L_b$ but $H_a\w e^a$ is of  
course
invariant as a consequence of (\ref{consis}). We may then write an  
anomaly
cancelling term on the world volume of the five-brane as,
\begin{equation}\label{}
L=\int_{W_6} B_a\w \chi_a(F).
\end{equation}
It should be observed that this is invariant under the $\L$ gauge
transformation of $B_a$ as a result of Stokes' theorem and the  
covariant
constancy of $\chi_a(F)$. In order to cancel
the anomaly we require, under  $SO(5)$ gauge transformation,  the  
variation
\begin{equation}\label{}
\d B_a=-\chi_a(\a, A) +(\a B)_a
\end{equation}
Now since the field strength $H_a\w e^a$ needs to be invariant we  
must replace
its relation to $B$ by,
\begin{equation}\label{H}
H_a=(d+A)_{ab}B_b+\chi_a^{C.S.}(A)
\end{equation}
so that it transforms covariantly under $\a$ gauge transformation.  
Using the
above descent formalism
we  get the Bianchi identity for this field,
\begin{equation}\label{}
 (DH)_a=F_{ab}\w B_b+\chi_a(F).
\end{equation}
{}From this  we have, using the covariant constancy of $e^a$ and the  
equation
(\ref{consis})
\begin{equation}\label{dH}
d(H_a\w e^a)=(DH)_a\w e^a=\chi_a(F)\w e^e.
\end{equation}
Integrating over any five cycle in $W_6$ we have the condition,
\begin{equation}\label{condition}
\int F^{ab}\w F^{cd}\w e^e\e_{abcde}=0.
\end{equation}
It appears then that we can cancel the anomaly coming from $p_2 (N)$  
provided
that we
require five-branes to satisfy the above restriction. However there  
remains the
problem of how to
reconcile the $H_a$ as defined above on $W_6$  with the four form  
field defined
on $M$. In
type IIA this was achieved as a result of the Thom isomorphism (see  
discussion
after (\ref{countertwo}).  In the present case the Bianchi identity  
for $K$ in
the presence of five-branes
is $dK\sim \d (M\rightarrow W_6)$. Picking up the relevant part of  
this
equation we have (writing also $\d  (M\rightarrow W_6)\sim \d_a\w  
e^a$) the
equation
$d (H_a\w e^a)|_W=(\d_a\w e^a)|_{W_6}$. We then need the result that  
the
`finite part of' the delta function can be chosen to be equal to the  
left hand
side of  (\ref{dH}).

In the above approach the anomaly normal bundle anomaly discovered in
\cite{ew4} appears to be treated differently from the other anomalies  
which are
cancelled by anomaly inflow from the bulk. Therefore let us take a  
somewhat
different approch which gives a unified treatment. This will also  
involve the
assumption  made
at the end of  the last paragraph and will turn out to be
closely related to the previous discussion.

It is convenient to
introduce the following redefinitions of our fields.
\begin{equation}\label{redef}
c_3={T_2C_3\over 2\pi},~c_6={T_5C_6\over 2\pi},~x_4=dc_3,~  
x_7=dc_6+\shalf
c_3\w x_4, ~b={T_2b_2\over 2\pi},~ h={T_2h_3\over 2\pi}.
\end{equation}
 In the presence of five branes we have $dx_4=\d_5$.  Since we have  
put
$2\k^2_{11}=1$ we find $T_2= (2\pi)^{2/3},~~T_5=(2\pi )^{1/3}$.
Also from the Hodge decomposition theorem we may write,  
$x_4=x_4'+\t_4$ where
the first term is a (unique) closed form (locally equal to $dc_3$)  
and the
second is a (unique) coexact form. So $d\t_4=\d_5$ and $d*\t=0$. It  
should also
be noted that with these normalizations the closed forms $x'_4$ and  
$x'_7=dc_6$
(locally) define integral classes. In particular $\int_{W_4} x_4\e  
{\cal
Z},~\int_{W_7} x'_7\e {\cal Z}$ (when $[\l ]$ is even integral).

We start again with the dual form of the M-theory action coupled to  
2- and
5-branes. It should be emphasized that we are using this action only  
as a guide
to
discovering the proper form of the five- and two-brane couplings in  
the usual
form of the action
and not as an end in itself.
 We have,
\begin{eqnarray}\label{k7}
I&=&-(2\pi)^{4/3}\shalf\int x_7\w *x_7+{2\pi\over 3}\int c_3\w x_4'\w  
x_4'+\nn
& &-2\pi\int_{W_3} c_3 +2\pi\int_{\pa W_3}b-{2\pi\over  
4}\int_{W_6}h\wedge *h -
2\pi \int_{W_6}c_6+{2\pi\over 2}\int_{W_6}c_3\wedge db+\ldots .
\end{eqnarray}
(The ellipses stand for pure gravity terms).
Writing (using the standard descent formalism)  
$\O_8=d\O_7^{C.S.},~{p_2\over
24}=d\o_7$ and
$\d\O_7^{C.S.}=d\O_6^1,~\d\o_7=d\o_6^1$, where $\d$ is a variation in  
the local
Lorentz group,
$SO(5,1)\times SO(5)$ in the first case and $SO(5)$ in the second,  
the anomaly
coming from (\ref{anomaly}) can be cancelled (as in the Green-Schwarz
mechanism) by assigning a local Lorentz transformation to $c_6$  
restricted to
$W_6$, i.e.
\begin{equation}\label{csixvar}
\d c_6 = -\O_6^1(TW_6\oplus N)+\o_6^1(N).
\end{equation}
Now again as in the Green-Schwarz mechanism the definition of the  
field
strength $x_7$ needs to be changed from (\ref{redef}) since the  
latter is no
longer gauge invariant. A neighbourhood of the five-brane world  
volume  $W_6$
may be defined  as a region $R$  of $M$ containing $W_6$ and bounded  
by  a
manifold $W_6\otimes S_4$. This is the region in which the structure  
group of
the tangent bundle of M effevtively becomes $SO(5,1)\otimes SO(5)$  
and its
transverse size may be regarded as the thickness of the five-brane.  
Let $\d_8$
be a gauge invariant closed eight form with compact support in R such  
that
$\d_8|_R\sim {p_2(N)\over 24}$. Explicitly we may take (following the  
previous
discussion) $\d_8 = \sum_{a=1}^5\d_{4 a}\w \chi^a(F)$, where the  
second factor
is defined after (\ref{p2n}) and the first factor is a four form with  
compact
support obtained by putting one index of the five-form $\d  
(M\rightarrow W_6)$
in the normal bundle. Then what we need is (as in the discussion  
after
(\ref{condition}) the generalization of the Thom isomorphism result  
used in
\cite{ew4}, namely
$\d_{4 a}|_{W_6}=\chi_a(F)$. Using the descent equations (see  
(\ref{des}) and
discussion)
we have
writing  $\d_8=d\t_7$ with $\t_7=\sum_a\d_a\w\chi^{a~C.S.}$ and
$\d\t_7=d\t_6^1$ for a gauge variation, where  
$\t_6^1=\sum_a\d_a\w\chi^a(A,\a
)$  with $\t_6^1|_{W_6}=\o_6^1$. Using these we may extend
(\ref{csixvar})  to $M$ and define the gauge invariant field,
\begin{equation}\label{x7}
x_7=dc_6-\shalf c_3\w x_4'-\O_7^{C.S.}(TM)+\t_7.
\end{equation}
Taking the exterior derivative,
\begin{equation}\label{dx7}
dx_7=-\shalf x_4'\w x_4'-\O_8+\d_8.
\end{equation}
Outside $R$ the last term is effectively zero and  this relation  
becomes the
usual one discussed in the literature $\cite{dlm}$.
Integrating over a closed\footnote{ Actually if we are in Minkowski  
space we
ought to be talking about configurations with compact support. It is
conceptually easier however to think in terms of compact spaces so  
that
strictly speaking the rest of the discussion is valid as stated only   
provided
it is understood that we have switched to a Euclidean metric on $M$.  
This also
implies that the topological terms are to be understood as having  
factors of
$i$.} eight manifold  we
get a restriction namely,

\begin{equation}\label{I8}
I_8=\int_{W_8}(-\shalf x'_4\w x'_4 -\O_8(TM) +\d_8)=0.
\end{equation}

The dual formulation of supergravity is however somewhat problematic.
It is not clear whether  a supersymmetric version exists. Also the  
relation to
the Horava-Witten theory is unclear.   Nevertheless the above
suggests a way to cancel the anomaly in the normal form of the  
theory.
So let us see how this  happens. Now there is no $c_6$ term so that  
it is
not clear how to write down a local six dimensional term that cancels  
the
anomalies. However it turns out that the relevant cancellation takes  
place by
anomaly inflow from the eleven dimensional bulk. Thus we write the  
following
expression for the complete (bosonic) action coupled to two and five  
branes:

\begin{eqnarray}\label{kfour'}
I&=&-\shalf (2\pi)^{2/3}\int_{M}x_4\w *x_4+{2\pi\over 3}\int_Mc_3\w  
x_4'\w
x_4'-2\pi\int_{M}x_4\w (\shalf x_4'\w c_3-\O_7+\t_7)\nn &
&-2\pi\int_{W_3}c_3+2\pi\int_{\pa W_3} b_2 -{2\pi\over  
4}\int_{W_6}h\w
*h-2\pi\int_{W_6}\shalf x_4'\w b_2,
\end{eqnarray}
with\footnote{ The topological terms over $M$ in this action (except  
for the
$\t$ term) appear to be essentially equivalent to those in  
\cite{ew4}. }
$x_4=x_4'+\t_4$ and $x_4'$ is the closed form which is locally equal  
to $dc_3$.
Note that this is the version of the normal form of the M-theory
action that is independent of any seven manifold that was promised at  
the end
of the last section. It should be noted that it reduces (ignoring the  
quantum
anomaly terms $\O,~\t$) to (\ref{kfour}) when we choose $\t_4
=-\t_4(M\rightarrow W_7^+)$ where the left hand side  is the step  
function
times delta function  restricting $M$ to a particular open seven  
manifold, that
was defined in the last section. It should also be noted that the  
above action
is dual to the action (\ref{k7}).

The gauge variation of the third term (under the transformations $\d
c_3=d\L$,and  local Lorentz transformations) gives, after using  
Stokes' theorem
and the modified Bianchi identity $dx_4=\d (M\rightarrow W_6)$,
\begin{equation}\label{}
-\int_M x_4\w (\shalf x_4'\w d\L_2-d\O_6^1+d\t_6^1)=\int_{W_6}  
(\shalf x_4'\w
\L_2-\O_6^1+\o_6^1).
\end{equation}
 The first term in the integrand cancels the anomaly coming from the  
last term
of (\ref{kfour'}) and  the second and third cancel the anomaly coming  
from
(\ref{anomaly}).

Now we need to argue that this term is well-defined.
Note that in the absence of five branes $x_4=x_4',~\t_7=0$ and  
(\ref{kfour'})
reduces to the standard
action of M-theory . In particular the topological term is (after  
using the
formula ${1\over 3} -{1\over 2} =-{1\over 6})$  exactly the one which  
was shown
to be well-defined in \cite{ew}. In the presence of five-branes  
however we need
to look at this issue again. In fact it does not seem possible to  
show this in
general. We will need to impose a topological restriction, namely  
that
$[\l]/2\equiv [p_1]/4\e \cal Z$.  The terms in question are (up to a  
factor
$2\pi$)
\begin{equation}\label{}
+{1\over 3}\int_Mc_3\w x_4'\w x_4'-\int_{M}x_4\w L_7
\end{equation}
where $L_7\equiv \shalf x_4'\w c_3-\O_7+\t_7$. Under the above  
topological
restriction it follows from the arguments of \cite{ew} that
the first term is well-defined, i.e. the integral over the closed  
12-manifold
${1\over 3}\int x_4'\w x_4'\w x_4'=0$. The second term may be  
rewritten by
using (\ref{x7}),
\begin{equation}\label{}
\int_M x_4\w x_7-\int_M x_4\w dc_6=\int_M x_4\w x_7  +\int_{W_6} c_6
\end{equation}
where we have used Stokes' theorem and the Bianchi identity $dx_4=\d$  
in the
last step. The
first term is clearly well-defined and the second term is  
well-defined, if
there is a well defined coupling of the six form field to the  
five-brane which
is the case when the Dirac quantization condition is satisfied. Thus  
a
well-defined action in the normal form exists provided that there is  
a
well-defined dual formualtion. Of course the definition of the  
five-brane
should also include
the restriction (\ref{condition}).

Finally let us check that our M-theory counter-term reduces to the  
one that is
given by
Witten \cite{ew4} in the IIA case. In that case the  counter term in  
question
is  $~ \int_{M_{10}} H_3\w \t_7 $ where now $\t_7\sim  
\d(M_{10}\rightarrow
W_6)\O^{C.S.}_{\chi} (F)$ (so that
$\d_8 \sim\d (M\rightarrow W_6)\chi (F))$ - with $\d_8|_R =\chi  
(F)^2\sim
p_2(N')$ - and $dH_3\sim \d (M\rightarrow W_6 )$. The counter term  
thus
becomes, $\sim \int_{W_6}H|_{W_6}\w \O^{C.S.}_{\chi}$ which is  
expression
(\ref{countertwo}),  the term given in \cite{ew4}.

\noindent{\bf Acknowledgments}\\
 I wish to thank Shyamoli Chaudhuri and Guoliang Yu  for discussions,  
and
Edward Witten for  very useful comments on section seven, and for   
explaining
some of the subtler points of \cite{ew4}.  I also wish to thank Mike  
Duff and
Peter West for discussions, and Eugen Cremmer, Bernard Julia, Hong  
Lu, and
Chris Pope  for pointing out an erroneous statement in an earlier  
version of
this paper.This work is partially supported by
the Department of Energy contract No. DE-FG02-91-ER-40672.

%%%%%%%%%%%%%%%%%%%%%%%%%%%%%%%%%%%%%%%%%%

%%%%%%%%%%%%%%%%%%%%%%%%%%%%%%%%%%%%%%%%%%%%%%%%
\end{document}